\documentclass[twocolumn]{aastex631}
\usepackage{multirow}
\usepackage{diagbox}
\usepackage{graphicx}

\graphicspath{{./}{Figure_FMR/}}

\makeatletter
\@ifundefined{longtable*}{}{%
  \expandafter\let\csname longtable*\endcsname\relax
  \expandafter\let\csname endlongtable*\endcsname\relax
}
\makeatother

\begin{document}

\title{Episodic Star Formation - II. The Origin of the Fundamental Metallicity Relation}

\author[0009-0005-9427-8525]{Xuelun Mei}
\affiliation{Department of Astronomy, Tsinghua University, Beijing 100084, China}

\author{Dandan Xu}
\affiliation{Department of Astronomy, Tsinghua University, Beijing 100084, China}

\author[0000-0002-8711-8970]{Cheng Li}
\affiliation{Department of Astronomy, Tsinghua University, Beijing 100084, China} 

\author[0009-0007-9365-9806]{Yuqian Gui}
\affiliation{Department of Astronomy, Tsinghua University, Beijing 100084, China}

\author[0009-0003-0166-0991]{Haoyi Wang}
\affiliation{Department of Astronomy, Tsinghua University, Beijing 100084, China}

\correspondingauthor{Xuelun Mei \& Dandan Xu}
\email{mxl25@mails.tsinghua.edu.cn}
\email{dandanxu@tsinghua.edu.cn}

\begin{abstract}
The Fundamental Metallicity Relation (FMR) extends the mass-metallicity relation (MZR) by including star formation rate (SFR) as a third parameter. At fixed stellar mass, galaxies with higher SFRs tend to have lower gas-phase metallicities.
We investigate the physical origin of the FMR in the context of episodic star formation (Gui et al. 2026, paper I of this series) by tracing the evolution histories of 8,979 central star-forming galaxies selected at $z=0$ in TNG100 back to $z=1$. We find that the overall MZR scatter at $z=0$ ($\sim0.11$ dex) comprises both long-lived differences among galaxies ($\sim0.09$ dex) and temporal fluctuations within individual galaxies ($\sim 0.07$ dex). 
The galaxy-to-galaxy component shows an anticorrelation between SFR and gas-phase metallicity, as shown by the observed FMR.
As for the consequence from temporal fluctuation, we track individual histories and stack the upward and downward SFR phases of a refined sample. 
Time-shift analysis and stacked histories show a well-behaved sequence in which the cold non-star-forming gas reservoir increases (decreases) first, followed by a decrease (increase) in gas-phase metallicity and then a rise (decline) in SFR. This ordering is consistent with a physical scenario where metal-poor gas first replenishes the cold reservoir and dilutes the gas before SFR rises to the peak, followed by feedback-driven gas heating/removal, resulting in a shift in metallicity from dilution to enrichment as SFR decreases towards the valley. This naturally explains the anticorrelation between the SFR and gas-phase metallicities as indicated by the observed FMR.
\end{abstract}


\section{Introduction} 
\label{sec:intro}
The gas-phase metallicity of a galaxy is a record of baryonic processes including gas accretion \citep{Forbes2014, DeLucia2020, vanLoon2021}, star formation \citep{Mannucci2010, EnciWang_2021}, metal production \citep{Zahid2014, JunYin2023}, feedback-driven outflows \citep{Tremonti2004, SimonLilly2013, Yang2024}, and recycling \citep{Muratov2017}. This makes it an important diagnostic of the baryon cycle in galaxy evolution. Observations have established numerous scaling relations linking metallicity to other galaxy properties, and these relations encode information about the assembly histories of galaxies (for a review, see \citealt{Maiolino_and_Mannucci2019}).

The most basic of these relations is the mass-metallicity relation, or MZR. The relation was first recognized in irregular and blue compact galaxies \citep{Lequeux1979} and later established with large local samples of star-forming galaxies \citep{Tremonti2004}. Galaxies with higher stellar masses tend to have higher gas-phase metallicities. This relation holds over a broad range of stellar masses and flattens at the high-mass end \citep{Tremonti2004, Curti2020}. There are several interpretations of the MZR.
On the one hand, feedback-driven outflows can remove a larger fraction of metals from low-mass galaxies because of their shallower gravitational potential wells \citep{Tremonti2004,ChengyuMa2024}. On the other hand, massive galaxies are generally at a more advanced stage of chemical evolution, consistent with downsizing \citep{Camps-Farina2021}, while low-mass galaxies tend to be more gas-rich, potentially reflecting ongoing accretion of metal-poor gas that dilutes the interstellar medium \citep{Lagos2016}. In addition, recycled gas can be more enriched in massive galaxies \citep{Brook2014}, and variations in the high-mass end of the stellar initial mass function may also shape the MZR by altering metal yields \citep{Koppen2007}.
The MZR is also observed at high redshift, with growing evidence that it extends to $z\sim10$ \citep{Nakajima2023highz, Curti2024highz, Sarkar2025highz}. However, at fixed stellar mass, high-redshift galaxies have lower metallicities than galaxies in the local Universe. This evolution has been linked to higher gas fractions \citep{Torrey2019} and more efficient metal loss through outflows \citep{Erb2006, Bassini2024}. 

At fixed stellar mass, the MZR shows substantial scatter that cannot be fully explained by measurement uncertainties. Further studies showed that part of this scatter is related to SFR. At fixed stellar mass, galaxies with higher SFRs tend to have lower gas-phase metallicities. This secondary dependence led to the proposal of the Fundamental Metallicity Relation (FMR), which links stellar mass, SFR, and gas-phase metallicity \citep{LaraLopez2010, Mannucci2010}. \citet{Mannucci2010} further suggested that the local FMR remains unchanged up to $z\sim2.5$. However, recent simulation studies challenged the idea of a universal, redshift-invariant FMR. Based on the Illustris, IllustrisTNG, EAGLE, and SIMBA simulations, \citet{Garcia2024FMR, Garcia2025FMRII} showed that the relation between MZR residuals and SFR can evolve with redshift.  JWST observations also find that galaxies up to $z\sim8$ are consistent with the local FMR of \citet{Andrews_and_Martini2013}, while those at $z>8$ are systematically more metal-poor \citep{Sarkar2025highz}. However, such comparisons across redshift are sensitive to metallicity calibrations and sample selection \citep{Pistis2024}. At subgalactic scale, integral field spectroscopy (IFS) surveys, such as MaNGA \citep{MaNGA2015} and CALIFA \citep{CALIFA2012}, have enabled studies of the spatially resolved FMR (rFMR), which links local gas metallicity, stellar mass surface density, and SFR surface density. However, whether metallicity exhibits a secondary dependence on SFR surface density at fixed stellar mass surface density remains debated \citep{Sanchez2013,Teklu2020,Baker2023rFMR}.

Analytic and semi-analytic gas-regulator models (e.g., \citealt{Eqmodel2012, SimonLilly2013}) provide key insight into connecting the MZR and FMR with the physical processes that regulate galaxy evolution. The basic premise these models rest on is that star-forming galaxies evolve slowly around a quasi-steady state, in which gas inflow, star formation, and outflows reach a balance. Departures from this equilibrium can arise from stochastic gas accretion. On shorter timescales, variations in gas supply produce episodic changes in gas content and metallicity relative to this baseline and can naturally produce an anticorrelation between SFR (or gas mass) and metallicity at fixed stellar mass \citep{Forbes2014, DeLucia2020, vanLoon2021, EnciWang_2021}. However, differences in the long-term equilibrium state can leave persistent offsets from one galaxy to another.

These analytic models necessarily simplify some relevant physical processes. Cosmological hydrodynamical simulations provide a complementary approach to interpreting the scatter of the MZR and the FMR by tracing gas accretion, star formation, feedback, and chemical enrichment in galaxies \citep{Torrey2018, Torrey2019, vanLoon2021, Garcia2024FMR, Garcia2025FMRII, Sharda2026}. Using Illustris and IllustrisTNG, \citet{Torrey2018} found that offsets from the star-forming main sequence and the MZR vary over similar timescales and are often anti-correlated, producing a pronounced FMR. Combining MaNGA observations with TNG50, \citet{ChengyuMa2024} found that galaxy size, which traces gravitational potential, is a stronger population-level predictor of metallicity than SFR, while SFR may modulate metallicity over time. These results suggest that the FMR may reflect both temporal fluctuations within individual galaxies and secular differences among galaxies. However, previous studies have neither quantitatively investigated the contributions of short-time and long-term behaviors to the observed MZR and FMR, nor delved into the detailed connection between episodic gas cycling (and star formation) and variations in gas metallicity.
 
This paper is the second in our episodic star-formation series. Using TNG100, \citet{SenWang_2022} showed that the large-scale environment regulates gas supply by modulating the angular momentum of the circumgalactic medium (CGM), producing episodic star formation in present-day star-forming galaxies. Building on this picture, Paper I \citep{Gui2026PaperI} provided a systematic overview of this episodic behavior. It also found that temporal fluctuations within galaxies and persistent differences among galaxies together account for the SFMS (star-forming main sequence) scatter. Here, we apply the same episodic framework to investigate the origin of the MZR scatter and the FMR. We focus on two questions. First, we quantify the relative contributions of secular differences among galaxies and temporal fluctuations within individual evolutionary histories to the MZR scatter. Second, we determine the temporal order of changes in gas supply, metallicity, and SFR during individual star-formation episodes. To address these questions, we select central star-forming galaxies at $z=0$ in TNG100, trace their main progenitors back to $z=1$, and quantify the population and temporal components of the MZR and SFMS scatter. We then stack the upward and downward SFR phases of galaxies with pronounced episodic star-formation histories (SFHs). We find that the present-day MZR scatter of $\sim0.11$ dex includes contributions from long-lived differences among galaxies ($\sim0.09$ dex) and temporal fluctuations within individual galaxies ($\sim0.07$ dex). During individual star-formation episodes, the cold non-star-forming gas reservoir changes first, followed by metallicity and then SFR.

The paper is organized as follows. Section~\ref{sec:method} describes the simulation, sample selection, definitions of the MZR and SFMS residuals, and stacking procedure. Section~\ref{sec:results_scatter} characterizes the MZR scatter and its relation to SFMS residuals. Section~\ref{sec:results_ESFandFMR_individual} presents the evolution tracks of individual galaxies in the MZR plane, and Section~\ref{sec:results_ESFandFMR_stacked} measures the relative time shifts among cold gas, metallicity, and SFR and presents the stacked evolution of gas mass, metallicity and oxygen mass during episodic star formation. Section~\ref{sec:conclusions} summarizes the main results and discusses their implications. Throughout the work, we adopt the flat $\Lambda$CDM cosmology used by IllustrisTNG, which is consistent with the Planck results \citep{Planck_Collaboration2016}. The relevant parameters are $\Omega_{\rm m}=0.3089$, $\Omega_{\rm b}=0.0486$, $\Omega_{\Lambda}=0.6911$, and $h\equiv H_0/(100\,{\rm km\,s^{-1}\,Mpc^{-1}})=0.6774$.
 
\section{Methodology} \label{sec:method}

\subsection{The TNG100 simulation}
\label{sec:method-TNG100}
IllustrisTNG (The Next Generation, hereafter TNG; \citealt{TNGCite02, TNGCite03, TNG50_02, TNGCite04, TNGCite05, TNGCite01, TNG50_01}) is a suite of cosmological, gravo-magnetohydrodynamical simulations run with the moving-mesh code \texttt{AREPO} \citep{AREPO}. The TNG consists of three volumes (TNG50, TNG100, and TNG300) with different resolutions and box sizes. In this work, we use TNG100, which provides a balance between volume and resolution. The TNG100 volume has a side length of $110.7\ {\rm Mpc}$, a baryonic mass resolution of $1.4\times10^6\,{\rm M_\odot}$, and a dark-matter particle mass resolution of $7.5\times10^6\,{\rm M_\odot}$. Dark-matter halos and their bound substructures are identified with the \texttt{SUBFIND} algorithm \citep{subfind_1,subfind_2}. We use the public TNG data products and merger trees released by the TNG Collaboration\footnote{\url{http://www.tng-project.org/data/}} \citep{TNGDataRelease}.

The TNG model follows the production and redistribution of metals self-consistently with stellar evolution. Dense gas cells ($n_{\rm H}\gtrsim 0.1\,{\rm cm}^{-3}$) are converted into star particles, each of which represents a single-age stellar population with a \citet{Chabrier2003} initial mass function and inherits the metallicity of its parent gas. 
As the stellar population evolves, mass and metals are returned to the surrounding interstellar medium (ISM) through three main enrichment channels: asymptotic giant branch (AGB) stars, core-collapse supernovae (SNII), and Type Ia supernovae (SNIa). The returned metals are deposited into neighboring gas cells and subsequently advected with the gas flow. Stellar-feedback-driven galactic winds further redistribute metals away from their sites of production \citep{TNGmodel}. The simulation explicitly tracks the abundances of H, He, C, N, O, Ne, Mg, Si, and Fe, in addition to the total metal mass.
In this work, we characterize the gas-phase metallicity using the oxygen abundance,
\begin{equation}
12+\log({\rm O/H}) =
12+\log_{10}\left(\frac{X_{\rm O}}{16X_{\rm H}}\right),
\label{eq:oh_def}
\end{equation}
where $X_{\rm O}$ and $X_{\rm H}$ are the oxygen and hydrogen mass fractions of the gas, respectively. And we use mass weighted gas-phase metallicity.

\subsection{Sample selection - I: Star-forming galaxies for evaluating the MZR and its scatter}
\label{sec:method-galaxysample}
We first select all central galaxies at $z=0$ with $10^9\,{\rm M_\odot}\leqslant M_{*,2R_{\rm hsm}}\leqslant 10^{11.5}\,{\rm M_\odot}$. Unless otherwise stated, stellar mass $M_*$, gas mass $M_{\rm gas}$, SFR, and metallicity $Z$ (characterized by oxygen abundance given by Eq.~(\ref{eq:oh_def})) are all measured within twice the stellar half-mass radius, $2R_{\rm hsm}$. The star-forming main sequence (SFMS) is determined iteratively in the $\log{\rm SFR}$--$\log\rm M_*$ plane using the galaxies in the mass range $10^9\,{\rm M_\odot}\leqslant M_{*}\leqslant 10^{10.5}\,{\rm M_\odot}$, where the main sequence can be considered linear \citep{Donnari2019}. Each time we fit a linear relation, remove galaxies more than 1 dex below the fitted line, and repeat the fit until convergence. The fitted SFMS is given by:
\begin{equation}
\log\left(\frac{\rm SFR}{\rm M_\odot\,Gyr^{-1}}\right)
=0.81\log\left(\frac{M_*}{\rm M_\odot}\right)+0.80.
\label{eq:sfms_z0}
\end{equation}
We note that both observations and cosmological simulations find evidence for a downward bend of the SFMS at the high-mass end \citep{Whitaker2014, Schreiber2015, Donnari2019}, although the strength of the bending depends on the adopted selection of star-forming galaxies. 

\begin{figure}[ht!]
    \centering
    \includegraphics[width=\linewidth]{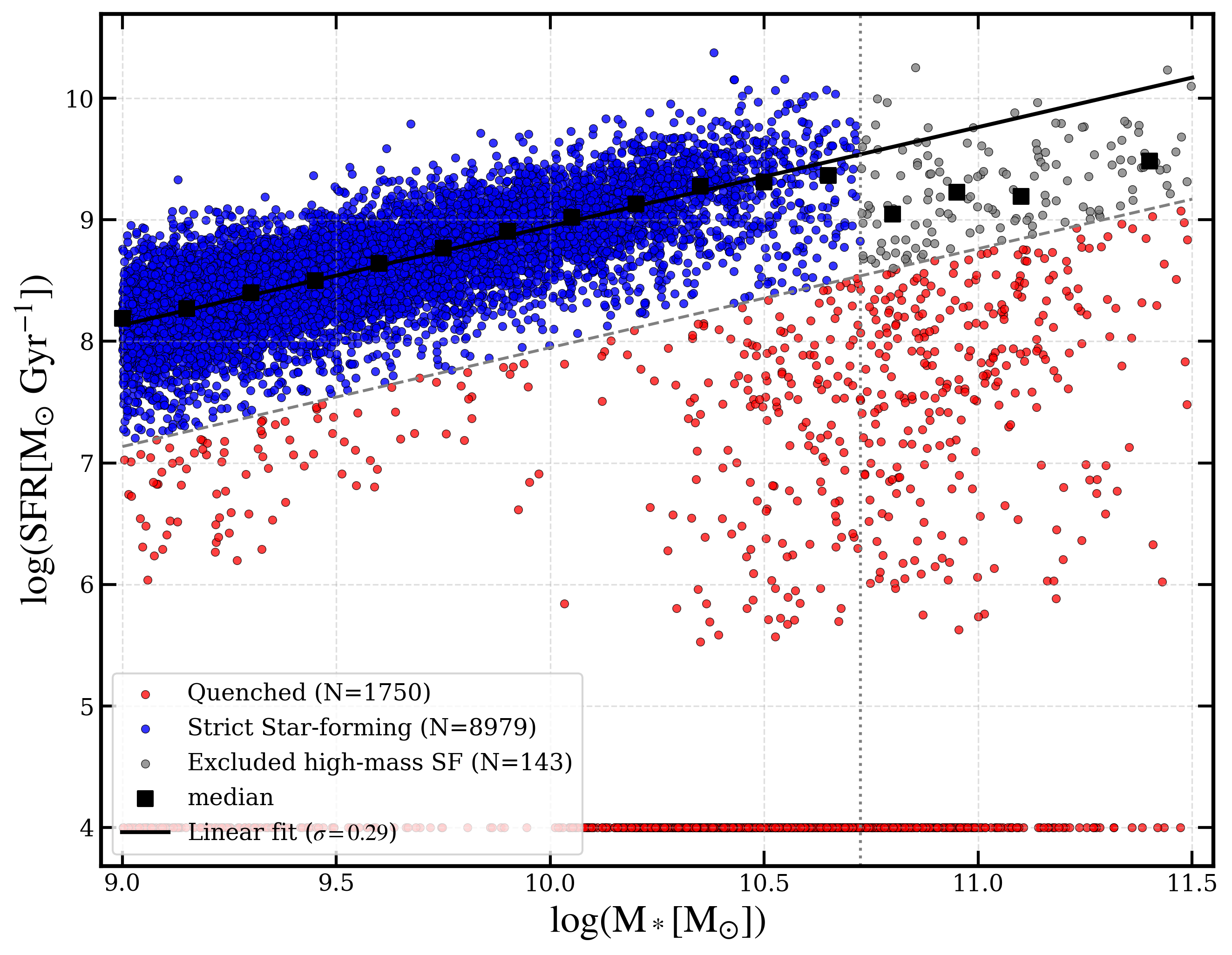}
    \caption{Selection of $z=0$ central star-forming galaxies in TNG100. The black solid line shows the iteratively fitted SFMS, and the gray dashed line shows the selection threshold 1 dex below the SFMS. The black squares show the median SFR of SF (star-forming) galaxies in each mass bin. The vertical gray dotted line shows the upper-mass boundary of strict SF sample. Blue, red, and gray points are the 8,979 central SF galaxies used in this work, quenched galaxies, and high-mass SF galaxies excluded by the strict stellar mass cut, respectively. The blue star-forming sample has a scatter of 0.29 dex about the SFMS.
    \label{fig:Sample_StrictSF}}
\end{figure}

In Fig.~\ref{fig:FMRatz0}, we present oxygen abundance versus stellar mass -- the MZR of the simulated central star-forming galaxies at $z=0$. While observed gas-phase MZR generally approaches a saturation metallicity at high stellar mass \citep{Tremonti2004, Zahid2014, Curti2020}, our TNG100 sample exhibits a mild downturn at the high-mass end. In addition, the anticorrelation between metallicity and SFR also becomes weaker toward the high-mass end. Existing studies have already shown that at $M_*\gtrsim10^{10.5}\,{\rm M_\odot}$, the anticorrelation nearly vanishes in SDSS and reverses sign in TNG and EAGLE, with the precise behavior depending on the adopted star-forming selection \citep{Mannucci2010, Torrey2019, Curti2020, Carnevale2026}.
These features motivate an additional upper stellar mass limit for our final  sample selection. For all 9,122 central star-forming galaxies that lie no more than 1 dex below the SFMS, we calculate the median SFR in stellar mass bins of 0.15 dex. We set the upper mass boundary at the first bin in which the median SFR falls more than 0.2 dex below the SFMS and exclude all galaxies above this boundary. At $z=0$, this criterion corresponds to $\log(M_*/{\rm M_\odot})\simeq10.725$ and results in a sample of 8,979 central star-forming galaxies (blue points in Fig.~\ref{fig:Sample_StrictSF}). 
We have checked that including these high-mass galaxies increases the scatter about the SFMS, but it does not change our scientific conclusions.

We then fit the MZR at $z=0$ of this sample with a third-order polynomial in $\log M_*$ and define $\Delta{\rm MZR}_{z=0}$ as the vertical offset in $12+\log({\rm O/H})$ from the fitted relation \citep{Yates2012, Curti2020}. To connect the present-day scatter to earlier evolution, we trace the main progenitors of each galaxy in our sample back to $z=1$. At each snapshot, we first re-identify the star-forming population using the same iterative SFMS procedure and fit the MZR at that epoch.
Following Paper~I \citep{Gui2026PaperI}, we characterize the scatter of the MZR (SFMS) with three quantities: (1) $\Delta{\rm MZR}_{z=0}$ ($\Delta{\rm SFMS}_{z=0}$) measures the present-day offset from the respective median line, (2) the time-averaged offset, $\langle\Delta \rm MZR\rangle_{z\leqslant 1}$ ($\langle\Delta \rm SFMS\rangle_{z\leqslant 1}$), quantifies how much a galaxy remains above or below the median line since $z=1$, thereby tracing long-term galaxy-to-galaxy differences, and (3) $\sigma_{\rm MZR,ind}$ ($\sigma_{\rm SFMS,ind}$) quantifies the scatter of temporal fluctuations within individual galaxies about their smooth long-term trends. For the metallicity of a given galaxy's progenitor at redshift $z_j$, we define $Z_j\equiv12+\log({\rm O/H})_j$ and fit $Z_j$ as a linear function of $\log(1+z_j)$ for each galaxy since $z=1$ \citep{Magnelli2014}. Denoting the value predicted by this fit at $z_j$ by $\widehat{Z}_j$, we then define
\begin{equation}
\sigma_{\rm MZR,ind}=
\left[\frac{1}{N}\sum_{j=1}^{N}
\left(Z_j-\widehat{Z}_j\right)^2\right]^{1/2},
\label{eq:sigma_mzr_ind}
\end{equation}
where $N$ is the number of snapshots for the galaxy at $z\leqslant 1$.
We define $\sigma_{\rm SFMS,ind}$ similarly for the temporal fluctuations in $\log{\rm SFR}$. The distributions and correlations of these quantities are presented in Section~\ref{sec:results_scatter}.

\subsection{Sample selection - II: Typical episodic galaxies for studying temporal evolution via stacking analysis}
\label{sec:method-galaxystacking}

To investigate whether and how episodic SFHs contribute to the MZR scatter, we follow the method of Paper I and further construct a refined sample of galaxies with pronounced episodic star formation out of the MZR sample composed of 8,979 $z=0$ central star-forming galaxies. Here we recap the key procedure; for more details we refer the reader to \citet{Gui2026PaperI}. An effective algorithm is used to first search for all peaks and valleys in the time evolution of historical SFRs. Adjacent peak and valley pairs are then identified, and the relative changes in SFR, $\Delta\log[{\rm SFR}/({\rm M_\odot\,Gyr^{-1}})]$, are calculated. A galaxy is classified as a member of this episodic sample if its SFH since $z=1$ contains at least two upward (valley-to-peak) pairs or at least two downward (peak-to-valley) pairs, with each pair satisfying $\left|\Delta\log[{\rm SFR}/({\rm M_\odot\,Gyr^{-1}})]\right|\geqslant 0.5$ dex. This criterion yields a parent sample of 6,382 galaxies ($\sim 71\%$). Although the stellar-mass range and the definition of the star-forming sample differ from those in Paper~I \citep{Gui2026PaperI}, applying the same episodic criteria yields similar fraction of episodic galaxies ($\sim71\%$ here versus $\sim70\%$ in Paper~I). Because of the computational cost of calculating cold-gas properties, we randomly select 200 galaxies in each of the two stellar-mass bins, i.e., $9.5\leqslant \log(M_*/{\rm M_\odot})<10.0$ and $10.0\leqslant \log(M_*/{\rm M_\odot})<10.5$, approximately preserving the distribution of the parent sample in the stellar mass–SFR plane. We have checked despite that metallicity is not included in the selection, its distribution in the resulting sample is also consistent with that of the parent sample (Fig.~\ref{fig:sample_distribution}).

Following Paper I, we trace two categories of {\it cold} gas elements: (1) cold, non-star-forming gas, defined as gas cells with an effective temperature $T\leqslant  2\times10^4\,{\rm K}$ and a number density below the adopted star-formation threshold, $n_{\rm H}\sim0.1\,{\rm cm}^{-3}$; and (2) star-forming gas, defined as gas cells above this density threshold. Together, the two make up the total {\it cold}-gas budget available for star formation.
For the refined episodic galaxy sample, we trace their merger histories and track the evolution of SFR, cold non-star-forming gas mass $M_{\rm cold,non-SF}$, star-forming gas mass $M_{\rm SF}$, and total cold-gas mass $M_{\rm cold}(=M_{\rm SF}+M_{\rm cold,non-SF})$ within $2R_{\rm hsm}$. To calculate the gas-phase metallicity and metal mass, we make no more distinction between the two gas categories and instead compute the oxygen abundance (to characterize metallicity) and oxygen mass ($M_{\rm O}=X_{\rm O}M_{\rm gas}$, to characterize metal mass) of total gas elements within $2R_{\rm hsm}$. Within this radius, the above-defined {\it cold} gas dominates the gas budget, accounting for $\sim95\%$ of the total gas mass — with the star-forming gas ($\sim 53\%$) contributing more than the cold, non-SF gas ($\sim 42\%$) at $z=0$. It is worth noting that the relative contributions from the two gas components to the total gas budget differ markedly between SFR peaks and valleys. As reported by Paper I, star-forming gas may account for $\gtrsim 80\%$ of the total cold gas at SFR peaks, but $\lesssim 50\%$ at SFR valleys.

In order to extract the general time evolution of key galaxy properties and pin down possible causes and consequences from the sequence of events, we measure time shifts of other galaxy properties relative to SFR and stack their histories over the identified SFR valley-to-peak (upward) and peak-to-valley (downward) phases.
For time-shift measurements, we use the residuals of the quantities that remove the stellar-mass and redshift dependence. We fit the median $\log{\rm SFR}$ – $\log M_*$ relation with a linear function and the median of other properties with third-order polynomials at each snapshot. We then subtract its fitted relation to obtain the residual time series for each quantity since $z=1$. 
For each galaxy, we interpolated the residual SFR and the residual $q$ into uniformly spaced time sequences with a timestep of $0.1\,{\rm Gyr}$ using cubic interpolation. 
The time shift $\tau_q$ is the lag that maximizes the cross-correlation between the residuals of $q$ and SFR, but limited by $|\tau_q| \leqslant 3\,{\rm Gyr}$. For the time shift associated with metallicity, we reverse the sign of the residual to account for its anticorrelation with SFR. We note that a negative time shift ($\tau_q<0$) indicates that the variation in quantity $q$ precedes the corresponding SFR variation, while a positive time shift ($\tau_q>0$) indicates that $q$ responds after the change of SFR. By construction $\tau_{\rm SFR}=0$.

As different galaxies possess star-formation episodes of different lengths in time, we do not directly stack their original time sequence by lining up a single epoch in time. 
For a given time span associated with each SFR peak-valley pair, we normalize the time coordinate by the time between the peak and the valley in the pair:
\begin{equation}
x =\frac{t-t_{0}}{|t_{\rm peak}-t_{\rm valley}|},
\end{equation}
where the starting point $t_0=t_{\rm valley}$ for a valley-to-peak episode and $t_0=t_{\rm peak}$ for a peak-to-valley episode. We then stack the histories of SFR, metallicity, oxygen mass, and total cold-gas mass after subtracting the value of each quantity at the start of the corresponding SFR episode ($t=t_0$). This preserves the relative temporal variations within each episode.

We remind the reader that the quantitative timing of the events is encoded in the distributions of $\tau_q$, presented in Table~\ref{tab:time_shifts}. The stacking analysis described above is not intended to quantify the time sequence; rather, it provides a qualitative picture of the ordering of events and thereby helps identify plausible cause-and-effect relations within the chain of events. To this end, we stack the normalized time sequences of the key galaxy properties defined above. Specifically, relative changes within a given phase are pooled into normalized time bins of width 0.1, and the sequences are stacked separately for the rising (valley-to-peak) and falling (peak-to-valley) phases. 
The stacking plot is presented in Section~\ref{sec:results_ESFandFMR_stacked}.

\section{The MZR and the FMR of the simulated star-forming central galaxies}
\label{sec:results_scatter}

\begin{figure}[ht!]
    \centering
    \includegraphics[width=\linewidth]{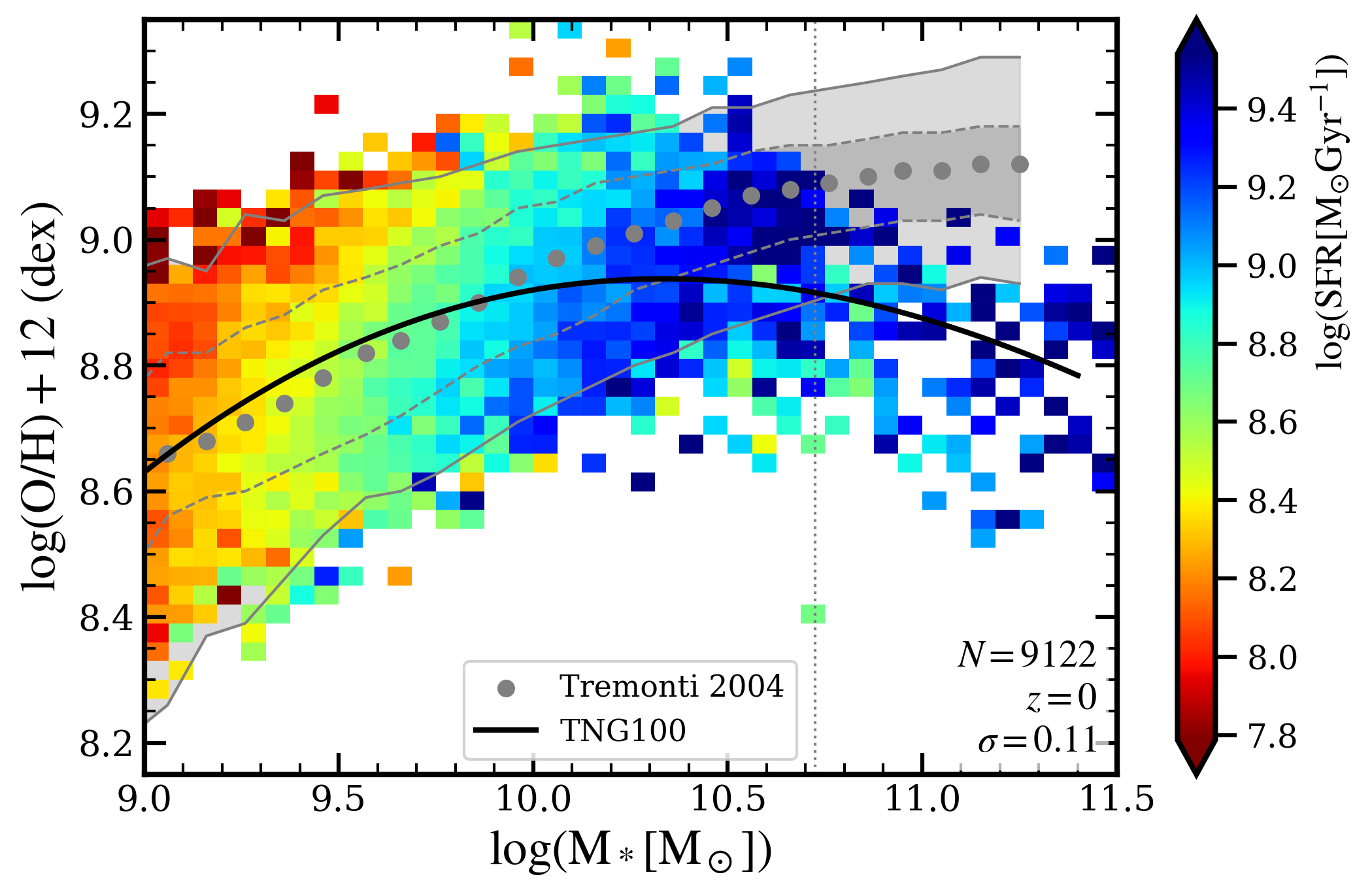}
    \caption{The gas-phase mass-metallicity relation of all central star-forming galaxies at $z=0$ in TNG100, color-coded by the mean SFR in bins of $\log M_*$ and $12+\log({\rm O/H})$. The black solid line is the median MZR. The vertical gray dotted line shows the same upper mass boundary of the selected star-forming galaxy sample as given by the blue points in Fig.~\ref{fig:Sample_StrictSF}. The gray points, along with the dark and light shaded regions, show the median relation and the central 68\% and 95\% distributions from \citet{Tremonti2004}.
    \label{fig:FMRatz0}}
\end{figure}

Fig.~\ref{fig:FMRatz0} shows the MZR of all central star-forming galaxies at $z=0$ in TNG100, with each bin color-coded by the mean SFR. The median metallicity increases with stellar mass before flattening at $M_*\sim10^{10.5}\,{\rm M_\odot}$, while the SFR color gradient shows that, at fixed stellar mass, galaxies with higher SFR tend to lie below the median MZR, i.e., possess lower gas-phase metallicities, essentially reflecting the FMR \citep{Mannucci2010, Torrey2019}. For comparison, we include observational data at $z=0$ from \citet{Tremonti2004}, for which the gray points show the median MZR, and the dark and light shaded regions show the central 68\% and 95\% distributions of SDSS star-forming galaxies. The TNG100 MZR broadly follows the observed shape at stellar masses below our upper mass cut for the central star-forming galaxy sample (see Section~\ref{sec:method-galaxysample}). At the high-mass end, however, the relation turns downward and lies systematically below the observational data, while the anticorrelation between SFR and metallicity nearly disappears. \citet{Torrey2019} found a similar flattening of the TNG100 MZR and attributed part of the underestimation of metallicity to contamination by nearly quenched galaxies.
AGN feedback also becomes increasingly important in this mass regime and can disturb the gas distribution and lower metal retention in TNG100 \citep{Weinberger2017,Weinberger2018,Torrey2019,Zinger2020}. Moreover, AGN feedback can change the SFR rapidly, whereas the low SFRs and gas accretion rates of these systems limit changes in their gas-phase metallicities. The resulting mismatch between the SFR and metallicity variability timescales weakens their anticorrelation \citep{Torrey2018}. \citet{Yang2024} found observational evidence for the same transition, with AGN activity becoming linked to MZR scatter above $M_*\sim10^{10.5}\,{\rm M_\odot}$.
Given this different behavior at the massive end, we use the strict star-forming galaxy sample defined in Section~\ref{sec:method-galaxysample}, bounded by the vertical dotted line in Fig.~\ref{fig:FMRatz0}, for the scatter analysis below.

\begin{figure*}[ht!]
    \centering
    \includegraphics[width=\linewidth]{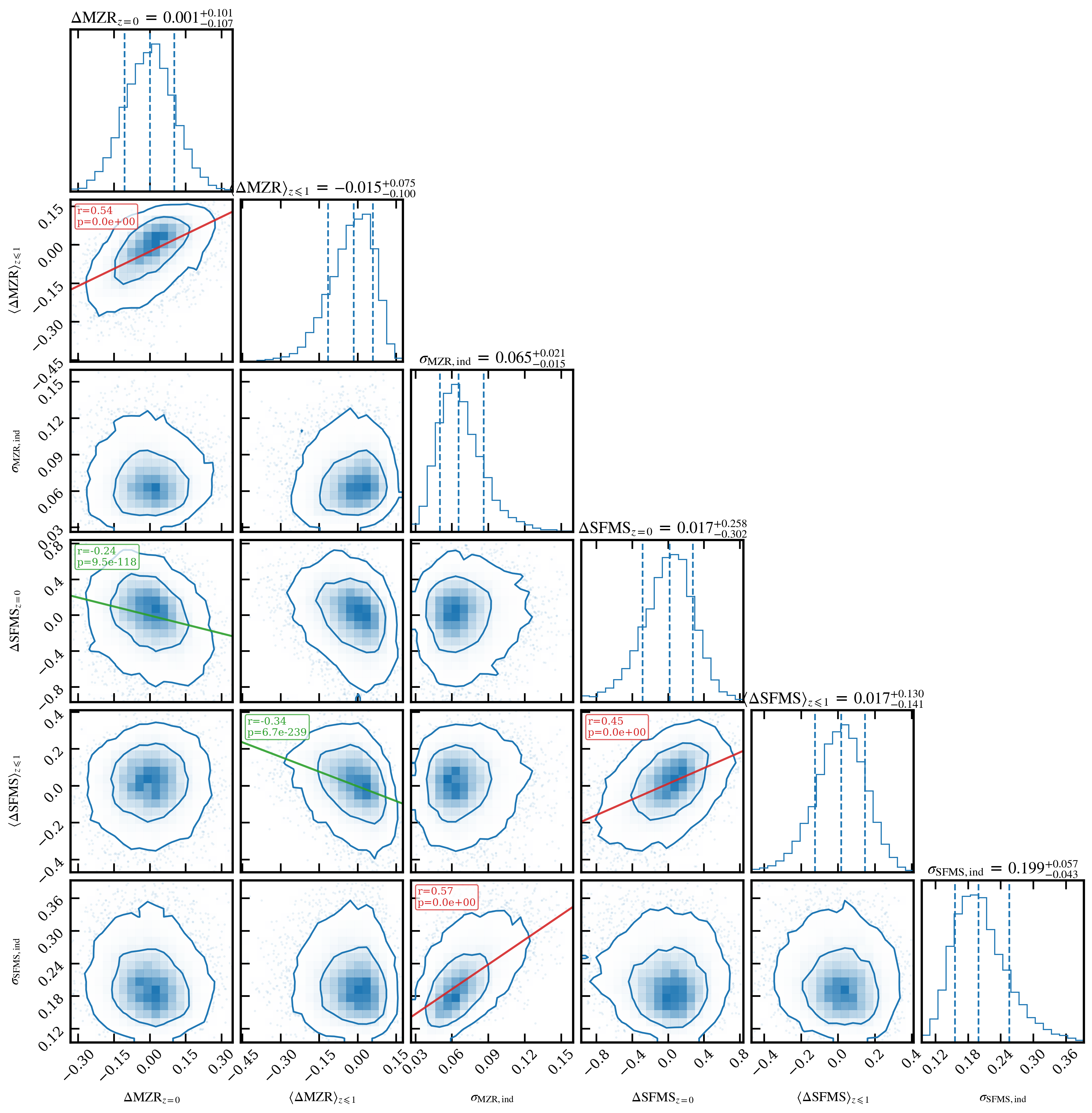}
    \caption{Distributions and pairwise relations of six measures of MZR and SFMS scatter: the offset at $z=0$ ($\Delta{\rm MZR}_{z=0}$ and $\Delta{\rm SFMS}_{z=0}$), the time-averaged offset since $z=1$ ($\langle\Delta{\rm MZR}\rangle_{z\leqslant 1}$ and $\langle\Delta{\rm SFMS}\rangle_{z\leqslant 1}$), and the temporal fluctuation of each galaxy since $z=1$ ($\sigma_{\rm MZR,ind}$ and $\sigma_{\rm SFMS,ind}$). All measurements are made within $2R_{\rm hsm}$. The diagonal panels show the distribution of each measure, with dashed lines marking the 16th, 50th, and 84th percentiles. The off-diagonal panels show the correlation between each pair, with contours enclosing 68\% and 95\% of the galaxies. Straight lines show linear fits for selected pairs. The annotations give the Pearson correlation coefficient $r$ and the corresponding two-sided $p$-value.
    \label{fig:corner_MZR}}
\end{figure*}

The gas-regulator models (e.g., \citealt{Eqmodel2012, SimonLilly2013, Forbes2014, EnciWang_2021}) predict that star-forming galaxies evolve around a slowly changing equilibrium balancing gas inflow, star formation, and outflows. Short-term variations in gas supply cause episodic fluctuations in gas content and metallicity, while long-term equilibrium differences create persistent offsets between galaxies.
In this work, following the decomposition of the SFMS scatter in Paper~I \citep{Gui2026PaperI}, we also quantitatively investigated how short-time and long-term behaviors contribute to the scatter of the MZR and the FMR. In Fig.~\ref{fig:corner_MZR}, the diagonal panels characterize these two contributions using the historically averaged MZR offset $\langle\Delta \rm MZR\rangle_{z\leqslant 1}$ and the averaged temporal fluctuation in individual galaxies $\sigma_{\rm MZR,ind}$, comparing to the present-day offset $\Delta{\rm MZR}_{z=0}$ (see Section~\ref{sec:method-galaxysample} for detailed definition).
At $z=0$, the 16th, 50th, and 84th percentiles of offset distribution are $\Delta{\rm MZR}_{z=0}=0.001_{-0.107}^{+0.101}$ dex, corresponding to an overall scatter of about $0.11$ dex. The historically averaged offsets since $z=1$ have a distribution of $\langle\Delta \rm MZR\rangle_{z\leqslant 1}=-0.015_{-0.100}^{+0.075}$ dex, with a characteristic spread of order $0.09$ dex -- typical long-term galaxy-to-galaxy diversity. The distribution of individual temporal fluctuations is $\sigma_{\rm MZR,ind}=0.065_{-0.015}^{+0.021}$ dex -- a typical temporal fluctuation of about $0.07$ dex. The present-day scatter of the MZR ($\sim0.11\rm \ dex$) therefore includes both long-lived diversity among galaxies ($\sim0.09\rm \ dex$) and fluctuation along individual evolutionary histories ($\sim0.07\rm \ dex$).

The off-diagonal panels of Fig.~\ref{fig:corner_MZR} show the pairwise relations among the six measures of MZR and SFMS scatter. We first focus on the correlation between the present-day and historically averaged offsets of the MZR and SFMS. The present-day MZR offset is positively correlated with its historical average since $z=1$ (the correlation coefficient $r\simeq0.54$), and the SFMS offset shows a similar correlation ($r\simeq0.45$). This means that a galaxy found above or below either median relation at $z=0$ is more likely to have remained on the same side of that relation, on average, since $z=1$. The present-day residual thus retains an imprint of each galaxy's averaged assembly history. This behavior is consistent with the long-lived MZR residuals found in EAGLE \citep{vanLoon2021} and with the long-timescale component of SFMS scatter found in simulations and observations \citep{NIHAO_Blank2021, Wan2025, Gui2026PaperI}.

We next focus on three panels in the lower left portion of Fig.~\ref{fig:corner_MZR}, which show the pairwise correlations between the corresponding present-day offset, historically averaged offset, and temporal fluctuations of the MZR and SFMS. At $z=0$, $\Delta{\rm MZR}_{z=0}$ anti-correlates with $\Delta{\rm SFMS}_{z=0}$ ($r\simeq-0.24$): galaxies with higher SFRs relative to the SFMS tend to have lower metallicities, as expected from the observed FMR. The modest correlation coefficient may arise partly because of the use of residuals across a large stellar mass range, since the strength of the secondary SFR dependence varies with mass \citep{Mannucci2010,Torrey2019}. The historically averaged offsets show a stronger anticorrelation ($r\simeq-0.34$), indicating that galaxy-to-galaxy diversity may provide a plausible source of the observed FMR. This long-term anticorrelation may reflect differences in gas supply, as continued accretion lowers the gas metallicity through dilution while sustaining star formation  \citep{Bothwell2013, Brown2018}. Differences in galaxy structure or gravitational potential may also contribute to this population-level trend \citep{ChengyuMa2024}. Finally, $\sigma_{\rm MZR,ind}$ correlates positively with $\sigma_{\rm SFMS,ind}$ ($r\simeq0.57$), so galaxies with larger SFR fluctuations also undergo larger metallicity fluctuations. 

\section{Evolution tracks of individual galaxies in the MZR plane}
\label{sec:results_ESFandFMR_individual}

The previous section showed that the historically averaged MZR and SFMS offsets of $z=0$ star-forming central galaxies exhibit clear anticorrelation, as observed for the present-day population. 
Such a long-term anticorrelation likely stems from differences in gas supply, or possibly from galaxy structure or gravitational potential \citep{Bothwell2013, Brown2018, ChengyuMa2024}. Apart from this long-lived galaxy-to-galaxy diversity which accounts for roughly 0.09 dex of the total 0.11 dex MZR scatter at $z=0$, there is also temporal variation in metallicity during the evolution of individual galaxies, contributing to 0.07 dex of the present-day MZR scatter. In this section, we investigate whether the expected SFR-metallicity anticorrelation also appears along evolution tracks of individual galaxies. 

\begin{figure*}[ht!]
	\centering
	\includegraphics[width=\linewidth]{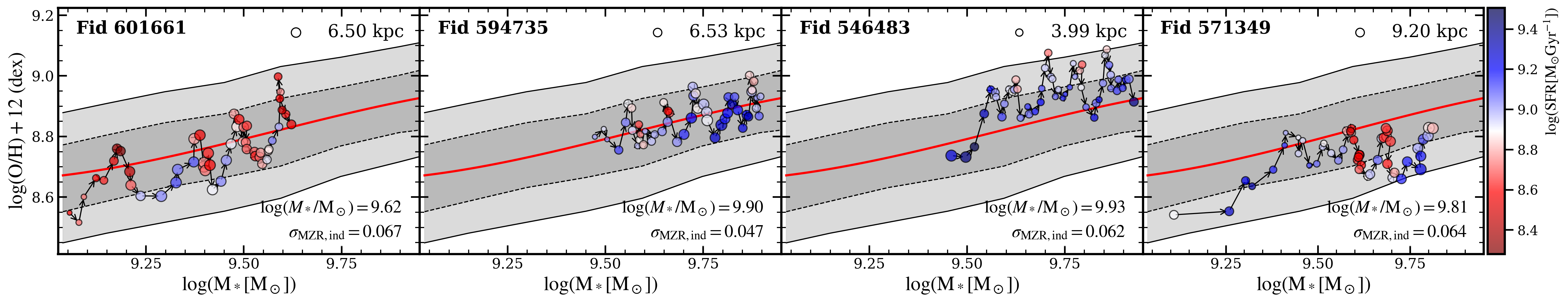}
	\includegraphics[width=\linewidth]{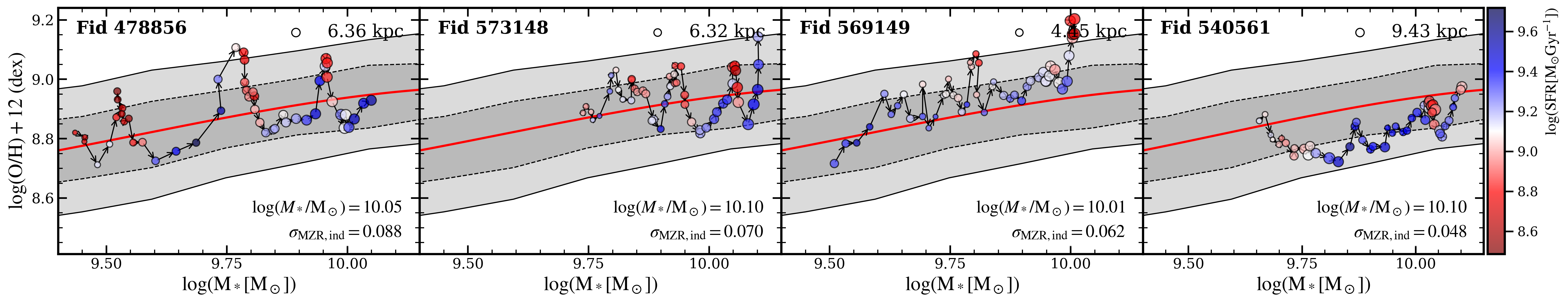}	
	\caption{Evolution tracks of individual episodic galaxies in the MZR plane. The upper and lower rows show galaxies selected from the low-mass $9.5\leqslant \log(M_*/{\rm M_\odot})<10.0$ and high-mass $10.0\leqslant \log(M_*/{\rm M_\odot})<10.5$ samples at $z=0$, respectively. Tracks in each panel show one galaxy from $z=1$ to $z=0$, with arrows pointing toward decreasing redshift. Points are color-coded by $\log[{\rm SFR}/({\rm M_\odot\,Gyr^{-1}})]$, with SFR increasing from red to blue. The point size scales with the physical size of $2R_{\rm hsm}$, with the open circle giving the median size in each panel. The red curve, together with the dark and light shaded regions show the median MZR and the central 68\% and 95\% ranges of the gas-phase metallicity distribution at fixed stellar mass for the strict star-forming sample at $z=0$. The value of $\sigma_{\rm MZR,ind}$ in each panel is the temporal fluctuation of metallicity, as defined in Equation~\ref{eq:sigma_mzr_ind}.}
	\label{fig:MZR_individual}
\end{figure*}

To examine the evolution of individual galaxies, we trace the main progenitors of typical episodic galaxies from $z=1$ to $z=0$. Fig.~\ref{fig:MZR_individual} shows the evolution tracks in the MZR plane of four low-mass galaxies in the upper row and four high-mass galaxies in the lower row. Both mass bins are defined by stellar mass at $z=0$ (Section~\ref{sec:method-galaxystacking}). Each point is one snapshot, with arrows indicating increasing cosmic time. The points are color-coded by $\log[{\rm SFR}/({\rm M_\odot\,Gyr^{-1}})]$, and the point size scales with $2R_{\rm hsm}$. The red curve shows the third-order fit to the $z=0$ MZR of the 8,979 strict star-forming sample, and the dark and light gray bands show its central 68\% and 95\% metallicity distributions. The median MZR is slightly steeper at $z=1$ than at $z=0$, and the scatter remains nearly unchanged. We therefore adopt the MZR at $z=0$ as a common visual reference for the evolution tracks. The value of $\sigma_{\rm MZR,ind}$ given in each panel quantifies the individual fluctuation of metallicity.

As can be seen, as a galaxy evolves, its stellar mass generally increases with time, whereas gas-phase metallicity fluctuates. The metallicity oscillates around its long-term enrichment trend, which evolves consistently to the median MZR at $z=0$. These fluctuations also show a similar dependence on SFR. Along a given track, phases with higher metallicity usually have lower SFR (redder), while phases with lower metallicity tend to have higher SFR (bluer). This result suggests that the SFR--metallicity anticorrelation is present not only across different galaxies but also within the evolutionary histories of individual galaxies.

In Fig.~\ref{fig:MZR_individual}, the tracks also reveal the galaxy-to-galaxy differences. In both mass bins, the galaxies in the third column lie predominantly above the median MZR (red curve) and have relatively small median values of $2R_{\rm hsm}$ (3.99 and 4.45 kpc), and those in the fourth column lie predominantly below the median and are more extended (9.20 and 9.43 kpc). Such systematic differences can be explained by \citet{ChengyuMa2024}, who adopted $M_*/R_{\rm e}^{\beta}$ to characterize the depth of the gravitational potential. They found that at fixed stellar mass, a smaller radius implies a deeper gravitational potential, which can improve metal retention and thus explain higher metallicity. Despite galaxy-to-galaxy differences, both the temporal metallicity variation and the SFR–metallicity anticorrelation along the evolutionary track are universal among individual galaxies. The metallicity evolution seen here is closely related to the picture in \citet{Torrey2019}, where individual galaxies show non-monotonic tracks also in the metallicity--ISM mass space.  
In the next section, we use the stacked time sequence to perform a detailed timeline analysis, revealing how gas mass, metallicity, and SFR vary and interconnect over a star-formation cycle.

\section{Stacked evolution during episodic star formation}
\label{sec:results_ESFandFMR_stacked}

In Paper I of this series, \citet{Gui2026PaperI} statistically investigated the temporal variation in the cold gas content during episodic star formation. They found that the cold non-SF gas variations precede the SFR variations, and the star-forming gas varies synchronously with SFR. The total cold gas, as predominated by the star-forming gas mass, closely tracks the SFR cycle, with a slight time lead. They interpreted this sequence as a cycle of gas supply, compaction, feedback and gas depletion. In this scenario, freshly accreted gas first builds the cold non-SF gas reservoir, which is subsequently compressed above the star-forming density threshold, feeding the SF gas pool, raising the SFR. After the SFR peak, feedback reduces the total cold gas budget and the SFR declines accordingly. As star formation and feedback weaken, the cold non-SF reservoir rebuilds and fuels the next episode. Building on this gas cycling scenario, we extend the analysis to the co-evolving temporal variation of gas-phase metallicity.

\begin{deluxetable}{l|cc}[h!]
    \tablewidth{0pt}
    \tablecaption{Median time shifts of gas masses and gas-phase metallicity relative to SFR and median durations of the upward and downward SFR phases, measured within $2R_{\rm hsm}$.\label{tab:time_shifts}}
    \tablehead{
    \multicolumn{1}{c|}{Quantity} &
    \colhead{Low mass} & \colhead{High mass}}
    \startdata
    \hline
    \multicolumn{3}{c}{Time shift $\tau_q$ (Gyr)}\\
    \hline
    $\Delta\log{\rm SFR}$ (reference) & $0.00$ (fixed) & $0.00$ (fixed)\\
    $\Delta\log M_{\rm cold,non-SF}$ & $-0.65_{-0.10}^{+0.10}$ & $-0.80_{-0.07}^{+0.11}$\\
    $\Delta\log M_{\rm cold}$ & $0.00_{-0.01}^{+0.00}$ & $0.00_{-0.01}^{+0.00}$\\
    $-\Delta[12+\log({\rm O/H})]$ & $-0.20_{-0.02}^{+0.01}$ & $-0.11_{-0.02}^{+0.00}$\\
    \hline
    \multicolumn{3}{c}{Time interval of SFR upward and downward pairs (Gyr)}\\
    \hline
    $T_{\rm upward}$ & $1.26_{-0.03}^{+0.05}$ & $1.24_{-0.03}^{+0.05}$\\
    $T_{\rm downward}$ & $1.32_{-0.03}^{+0.08}$ & $1.34_{-0.03}^{+0.08}$
    \enddata
    \tablecomments{Low and high mass correspond to $9.5\leqslant\log(M_*/{\rm M_\odot})<10.0$ and $10.0\leqslant\log(M_*/{\rm M_\odot})<10.5$, respectively. A negative $\tau_q$ indicates that $q$ precedes SFR. The metallicity residual is sign-reversed when calculating the cross-correlation. Subscripts and superscripts give the uncertainties on the median values, derived from the 16th and 84th percentiles around the median divided by $\sqrt{N}$. Here, $N$ is the number of valid galaxies for time shifts and the number of SFR episodes for phase durations. For time shifts, $N=200$ galaxies in each mass bin. The numbers of upward SFR episodes are 385 and 366, and the numbers of downward SFR episodes are 409 and 400, for the low- and high-mass bins, respectively.}
\end{deluxetable}

We first quantify the time shift $\tau_q$ of a given residual property $q$ relative to the temporal variation of SFR, using the cross-correlation procedure described in Section~\ref{sec:method-galaxystacking}. Table~\ref{tab:time_shifts} presents median $\tau_q$ for several key quantities. By construction, the time sequence of the SFR offset $\Delta\log{\rm SFR}$ has $\tau_{\rm SFR}=0$. Because SFR in the simulation arises essentially from star-forming gas, the residual star-forming gas mass $\Delta\log M_{\rm SF}$ thus also has $\tau_{M_{\rm SF}}=0$ by default. The table also presents the time shifts of the residual cold, non-SF gas mass $\Delta\log M_{\rm cold,non-SF}$ and the negative metallicity residual, $-\Delta[12+\log({\rm O/H})]$. In the bottom part of the table, we give the median durations of the upward and downward SFR phases. All quantities above are measured within $2R_{\rm hsm}$ for galaxies in both stellar-mass bins, with times given in unit of Gyr.  

Because the total cold‑gas budget is dominated by the star‑forming gas, the total cold-gas mass tracks the star-forming gas mass and the SFR closely. The non-star-forming cold gas, however, shows fairly significant median time leads of $0.65–0.80$ Gyr between the two mass bins, over a total median duration of $T_{\rm upward}\sim$1.26 Gyr for the rising phase and $T_{\rm downward}\sim$1.33 Gyr for the falling phase. 
In comparison, the negative metallicity variations have median time leads of $0.11–0.20$ Gyr between the two mass bins, which are several hundred million years behind the variation of the cold, non-star-forming gas mass, but only slightly ahead of the SFR variation. It is this much shorter shift between the negative metallicity variation and the SFR that makes the two sequences swing in almost opposite directions over the episodic cycles, producing a marked anticorrelation. 

We note the reader that the error on the quoted median has two parts. The first kind of uncertainty comes from the method itself. Deriving the time shift by cross-correlating two time sequences introduces an aliasing ambiguity: because the sequences are quasi-periodic, the cross-correlation shows secondary peaks at integer multiples of the period, so the same data are consistent with lags offset by one or more periods. Even though we verify that the peak we use is the primary one, this ambiguity cannot be fully removed, because the sequences are truncated to $z\leqslant 1$ and therefore span only a limited number of periods. 
The second part comes from galaxy-to-galaxy scatter: with only a finite number of galaxy tracks, the complex physical processes acting on individual galaxies add uncertainty to the median time shift.
For example, $M_{\rm SF}$ is the immediate fuel reservoir and is closely coupled to SFR. In contrast, $M_{\rm cold,non-SF}$ is an upstream reservoir regulated by gas accretion and cooling, compaction into the star-forming phase, and feedback-driven heating or removal. These competing channels weaken its direct coupling to SFR and also introduce uncertainty to the median time shift.

\begin{figure*}[ht!]
    \centering
    \includegraphics[width=\linewidth]{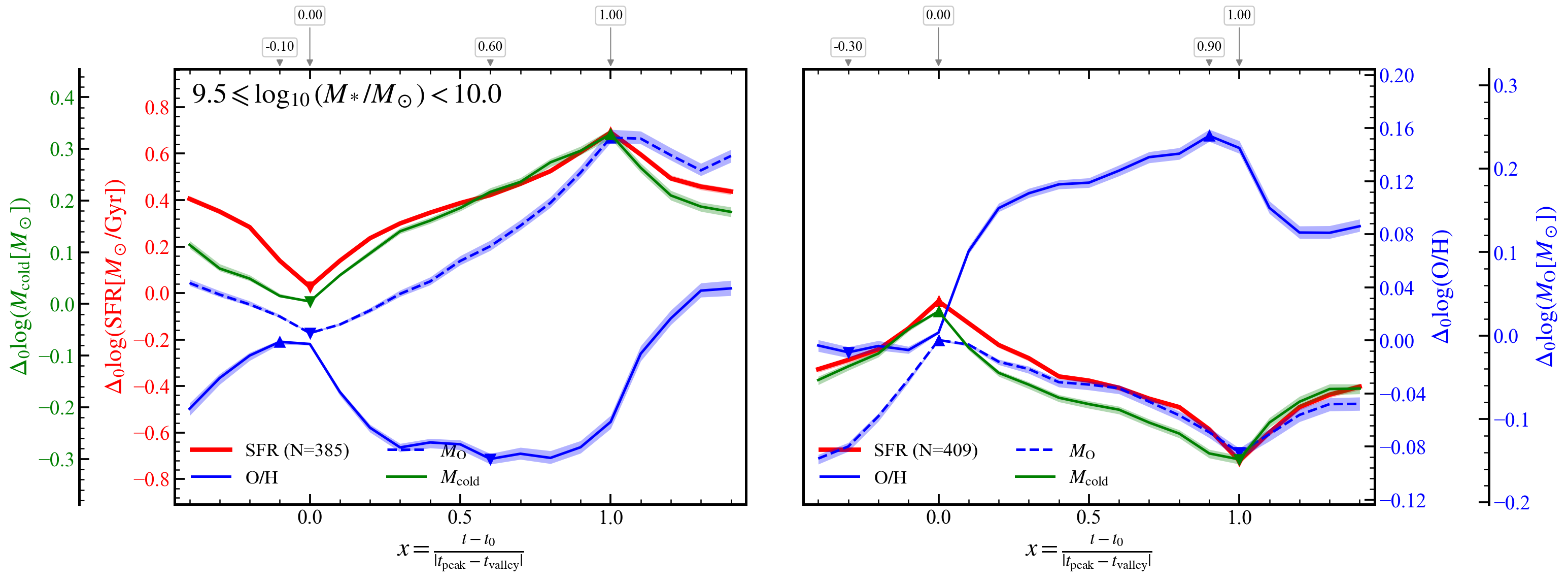}
    \includegraphics[width=\linewidth]{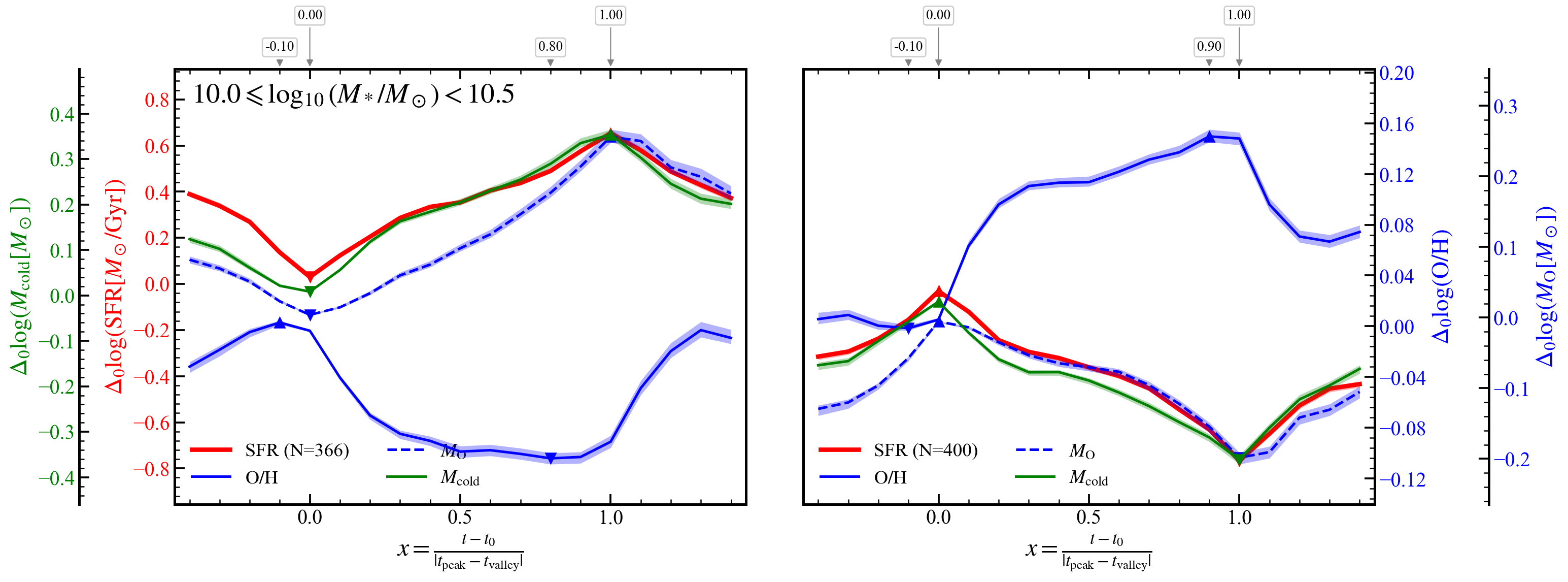} 
    \caption{Stacked relative evolution of SFR (thick red), metallicity (blue solid), gas-phase oxygen mass (blue dashed), and total cold gas mass (green) during upward (left) and downward (right) SFR phases. The upper and lower rows correspond to galaxies with $9.5\leqslant \log(M_*/{\rm M_\odot})<10.0$ and $10.0\leqslant \log(M_*/{\rm M_\odot})<10.5$, respectively. The time normalization procedure is described in Section~\ref{sec:method-galaxystacking}. The shaded regions represent the uncertainties in the median estimated from the 16th and 84th percentiles around the median, divided by $\sqrt{N_{\rm SFR}}$, where $N_{\rm SFR}$ is the number of SFR episodes indicated in each panel. Triangles mark the highest local maximum and lowest local minimum of the stacked curves, and their dimensionless $x$ coordinates are annotated above each panel.
    \label{fig:gasZ_separated}}
\end{figure*}

We further build a qualitative picture to more clearly present the event ordering and relative evolution. This is achieved by stacking the normalized time sequences for individual galaxies over time spans associated with the rising or falling SFR phases (see Section~\ref{sec:method-galaxystacking} for details). Before stacking, we subtract each quantity's value at the start of the corresponding SFR episode ($t=t_0$). We denote these changes by $\Delta_0\log Q\equiv\log Q(t)-\log Q(t_0)$, distinguishing them from the residuals relative to the fitted population relations. 
The resulting stacked curves for galaxies in the two mass bins are shown in Fig.~\ref{fig:gasZ_separated}.
Since the star-forming gas mass essentially follows the track of SFR, we do not show them in the same figure, but only present the relative evolution of total cold gas mass $\Delta_0\log M_{\rm cold}$ (green), metallicity $\Delta_0\log({\rm O/H})$ (blue solid) and gas-phase oxygen mass $\Delta_0\log M_{\rm O}$ (blue dashed), in comparison to $\Delta_0\log{\rm SFR}$ (thick red) during the episodic star formation. The left panels show the time sequences associated with the rising phase of SFR, where $x=0$ corresponds to the SFR valley, while the right for the falling phase, where $x=0$ corresponds to the SFR peak. On top of each panel, we label out the closest local valley/peak in the median sequence for each of the four quantities relative to $x=0$. We note that the cross-correlation method used to determine the optimal time shift neither assumes nor ensures that the periods of the compared time series are similar. The time-offset readouts therefore give the averaged peak–valley time interval for each investigated quantity.  

As can be seen, the stacked evolution is similar in the two mass bins. Across both columns and for both mass ranges, the upward (downward) variation in $\Delta_0 \log M_{\rm cold}$ and $\Delta_0 \log M_{\rm O}$ and the downward (upward) variation in $\Delta_0 \log({\rm O/H})$ have approximately the same duration as the median upward (downward) SFR phase. This agrees with \citet{Torrey2018}, who found that SFR and metallicity in TNG evolve on similar timescales that are well matched to the halo dynamical time. In both mass bins, we also see that $\Delta_0\log M_{\rm O}$ and $\Delta_0\log M_{\rm cold}$ rise and fall approximately in step with SFR. By contrast, $\Delta_0\log({\rm O/H})$ begins to decline shortly before SFR rises and to increase shortly before SFR falls.
The metallicity variation leads that of the SFR by only a small interval ($0.1{-}0.2$ Gyr versus a valley-to-peak time of $\sim 1.3$ Gyr; see Table~\ref{tab:time_shifts}). Consequently, throughout the rise (fall) of the SFR ($0 \leqslant x \leqslant 1$), the metallicity almost always moves in the opposite sense, falling (rising) from its peak (valley) to its valley (peak), and thus exhibits an anticorrelation with the SFR.      

We note that metallicity as approximated by the oxygen abundance $12+\log({\rm O/H})$ is essentially the ratio between the metal mass and the total gas mass. Within $2R_{\rm hsm}$ the total gas mass is overwhelmingly dominated by the sum of the cold, non-SF gas and the SF gas, together referred to as the cold gas ($\sim 95\%$). The temporal variation of the cold gas mass (green lines), which closely follows that of the SFR (red thick lines), is governed by both supply and removal processes. The cold gas budget is supplied by two sources: fresh cold gas accreted from external origins, and metal-polluted hot gas that is first transported to large radii and then cools and recycles back into the galactic disk. Cold gas is removed through two channels: dense compression to form stars, and heating to higher temperatures or expulsion beyond the evaluation aperture due to feedback. The cause of metal mass evolution is similar: the increase is due to star formation and feedback returning metals to the ISM, and gas flowing into the aperture, while the decrease is due to feedback expelling metals beyond the evaluation aperture.

In order to better understand the relative changes in oxygen mass, total cold-gas mass, and metallicity, we followed \citet{Garcia2025FMRII} and adopted a constant hydrogen-to-gas mass fraction of 0.76. Approximating the total gas mass by the cold-gas mass then gives $\Delta\log M_{\rm O}\simeq\Delta\log M_{\rm cold}+\Delta\log({\rm O/H})$. This relation provides an approximate guide to interpreting the relative changes in the stacked curves. During the SFR-rising phase, in both mass bins, the cold-gas mass $\Delta_0 \log M_{\rm cold}$ increases by $\sim0.35$ dex and the oxygen mass $\Delta_0 \log M_{\rm O}$ also increases by $\sim0.25$ dex, while the metallicity $\Delta_0 \log({\rm O/H})$ decreases by $\sim0.10$ dex. Thus, the metal mass grows along with the cold-gas reservoir, but by a smaller fractional amount, leading to a declining metallicity trend. This behavior is due to a substantial supply of low-metallicity gas, with dilution outweighing enrichment. During the SFR-declining phase, $\Delta_0 \log M_{\rm cold}$ decreases by $\sim0.30$ and $\sim0.35$ dex in the low- and high-mass bins, respectively, and the oxygen mass $\Delta_0 \log M_{\rm O}$ decreases by $\sim0.15$ and $\sim0.20$ dex. In both mass bins, the metallicity $\Delta_0 \log({\rm O/H})$ increases by $\sim0.15$ dex. The metal mass declines less rapidly than the cold-gas mass, resulting in a rising metallicity trend. In this phase, although metal production weakens, continued enrichment still dominates over dilution by fresh gas because the cold gas supply has become progressively weaker. The same qualitative behavior is also found when the analysis is repeated within a fixed $6\,{\rm kpc}$ aperture. These results suggest that, during episodic star formation, gas cycling couples fluctuations in SFR to the competition between metal dilution and enrichment, causing SFR and metallicity to vary on similar timescales while remaining nearly anti-correlated. 

\section{Summary and Discussion}
\label{sec:conclusions}

This is the second paper in our episodic star-formation series. Paper~I \citep{Gui2026PaperI} established an episodic picture for the recent evolution of present-day SFMS galaxies in TNG100. Here, we extend this episodic framework to investigate the origin of the MZR scatter and the FMR for star-forming galaxies in TNG100. We select 8,979 central star-forming galaxies at $z=0$ and follow their main progenitors back to $z=1$. We characterize the MZR scatter using the present-day offset, the historically averaged offset, and the temporal fluctuation of each galaxy. We then use a refined sample of 400 galaxies with pronounced episodic SFHs to measure the relative times and amplitudes of variations in SFR, gas-phase metallicity, metal mass, and the cold gas reservoir. Our main conclusions are as follows.

\begin{enumerate}
    \item TNG100 broadly reproduces the shape of the $z=0$ MZR and its secondary dependence on SFR below the high-mass end (Fig.~\ref{fig:FMRatz0}). At $M_*\gtrsim10^{10.5}\,{\rm M_\odot}$, however, the MZR turns downward relative to the SDSS relation and the SFR--metallicity anticorrelation weakens. We adopt a strict upper boundary of $\log(M_*/{\rm M_\odot})\simeq10.725$ for the main analysis. Within this sample, the present-day MZR has a scatter of $\sim0.11$ dex. The historically averaged offsets since $z=1$ give a typical population diversity of about $0.09$ dex, and the median individual temporal fluctuation is $\sigma_{\rm MZR,ind}\simeq0.07$ dex (Fig.~\ref{fig:corner_MZR}). These statistics show that both long-lived galaxy-to-galaxy diversity and temporal fluctuations within individual galaxies contribute to the observed MZR scatter.

    \item The FMR contains a long-term component associated with galaxy-to-galaxy diversity (Fig.~\ref{fig:corner_MZR}). The present-day MZR offset is correlated with its historical average since $z=1$ ($r\simeq0.54$), showing that a galaxy's current position relative to the MZR retains information about its long-term evolution. More importantly, the historically averaged MZR and SFMS offsets are anti-correlated ($r\simeq-0.34$). This long-term anticorrelation contributes to the FMR, possibly linked to differences in gas content and gravitational potential.  

    \item Individual episodic evolution also contributes to the FMR. Galaxies oscillate around their long-term MZR trends, with higher SFR phases tending to be more metal poor (Fig.~\ref{fig:MZR_individual}). The time-shift analysis shows that the cold non-SF gas reservoir leads SFR by a median of $\sim0.7\,{\rm Gyr}$. The anti-correlated metallicity variation leads SFR by only $0.20$ and $0.11\,{\rm Gyr}$ in the low- and high-mass bins, respectively (Table~\ref{tab:time_shifts}). The stacked histories further show that both the total cold-gas mass and metal mass increase during the SFR-rising phase and decrease during the SFR-declining phase (Fig.~\ref{fig:gasZ_separated}). This sequence suggests that the non-SF gas reservoir builds up first and dilutes the ISM during the upward SFR phase. Part of this gas then reaches the SF gas phase and raises the SFR. During the downward SFR phase, dilution weakens and enrichment becomes relatively more important.

\end{enumerate}

These results suggest that the FMR in TNG100 arises from the combined effects of long-term galaxy-to-galaxy diversity and episodic fluctuations within individual galaxies. However, the quantitative contributions of these two components to the MZR scatter and the FMR depend on the TNG galaxy-formation models. Stellar and AGN feedback regulate gas supply, recycling, and metal retention, while the unresolved multiphase ISM and finite resolution affect metal mixing. Comparisons with other simulations would help assess the robustness of these results to different galaxy-formation models. Extending this decomposition to high redshift will show how the relative contributions of the two components change and help us investigate whether the FMR evolves with redshift.

Finally, we note the reader that the results evaluated within $2R_{\rm hsm}$ may to some extent vary quantitatively depending on the aperture choice for evaluation. This is because, as expected from the two-branch star-formation picture presented in Paper~I \citep{Gui2026PaperI}, while the inner, higher-metallicity branch of star formation is sustained mainly by enriched gas in the central region, the outer branch, during each star-formation episode begins in the lower-metallicity galactic outskirts, inbreathing freshly arrived gas, and subsequently moves inward with gradually rising metallicity. On the one hand, this radial progressing in star formation and metal enrichment can potentially result in different time signatures for the variation sequences in SFR and metallicity at different radii. On the other hand, it is exactly such episodic gas cycling and star formation across the entire galaxy that regulate the temporal rhythm of metal enrichment. The reported anticorrelation between SFR and gas-phase metallicity thus may be more of a global relation. It is therefore highly interesting to investigate the radial dependence. In the third paper of this series (Mei et al. in prep), we investigate the detailed dynamical evolution of the gas components across a large radial range of the galaxy in the context of episodic star formation.

\section*{Acknowledgements}
We thank Profs. Stijn Wuyts and Enci Wang for their insightful discussions and constructive suggestions. This work was supported by the National Key Research and Development Program of China (grant No. 2022YFA1602903), the National Natural Science Foundation of China (grant No. 12433003), and the China Manned Space Project (No. CMS-CSST-2025-A10). We thank the IllustrisTNG Collaboration for making the simulation data publicly available and acknowledge the use of its JupyterLab environment to access and explore the TNG simulations.

\appendix

\section{Distributions of the Refined Episodic Sample}
\label{sec:appendix_sample_distribution}
\begin{figure*}[ht!]
   \centering
   \includegraphics[width=0.8\textwidth]{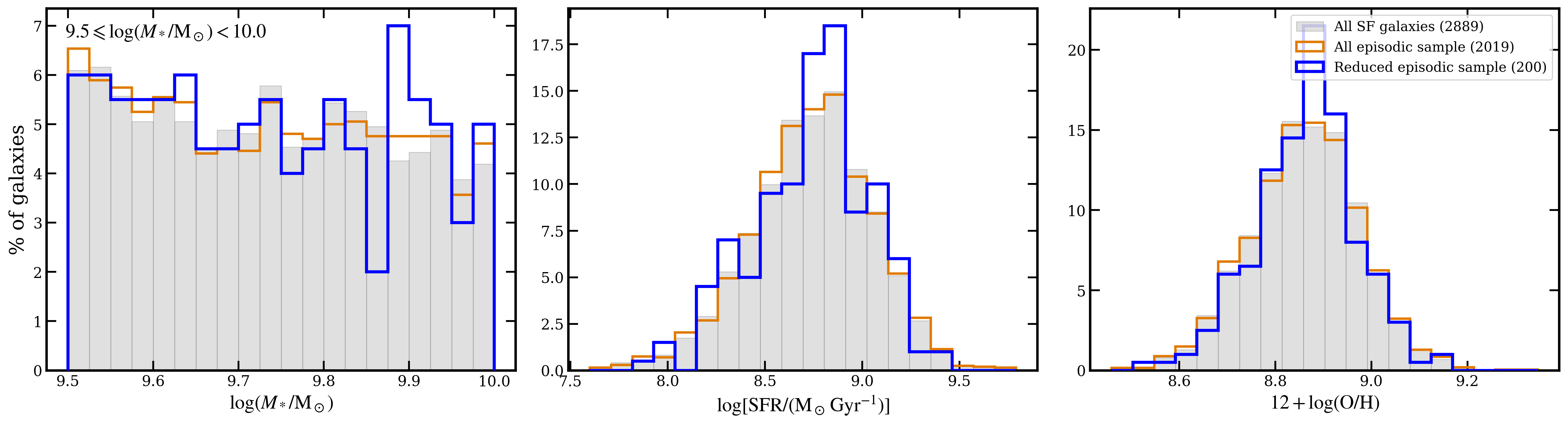}
   \includegraphics[width=0.8\textwidth]{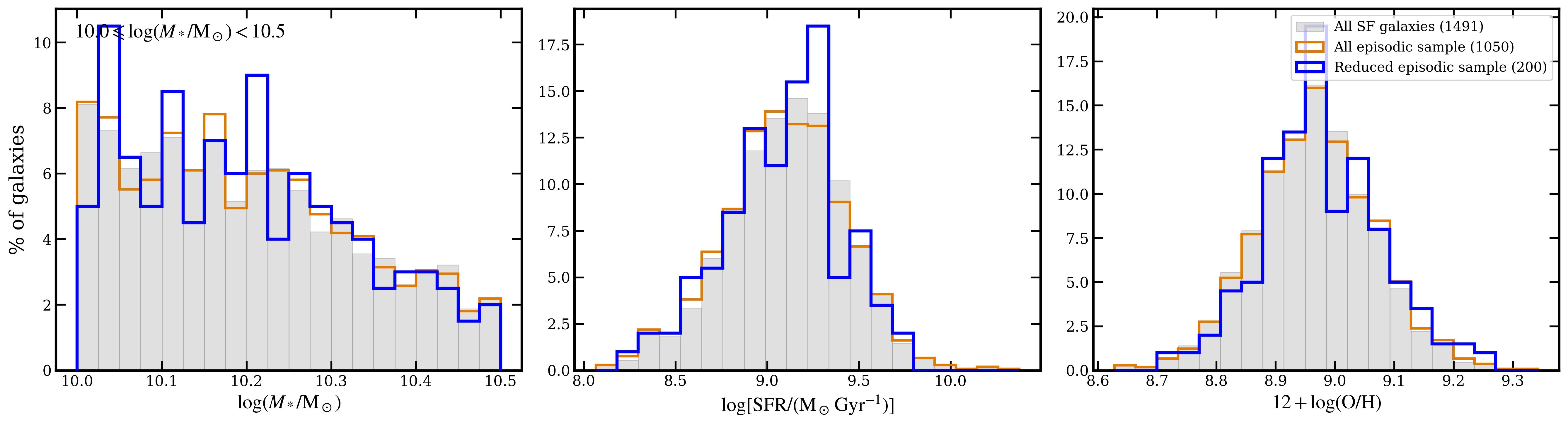}
   \caption{Distributions of stellar mass (left), SFR (middle), and gas-phase metallicity (right) for the refined episodic sample used in the stacking analysis. The upper and lower rows correspond to the stellar mass ranges $9.5\leqslant\log(M_*/{\rm M_\odot})<10.0$ and $10.0\leqslant\log(M_*/{\rm M_\odot})<10.5$, respectively. The gray filled histograms show all central star-forming galaxies, and the orange and blue step histograms show the parent and reduced episodic samples.}
   \label{fig:sample_distribution}
\end{figure*}

Fig.~\ref{fig:sample_distribution} compares the normalized distributions of stellar mass, SFR, and metallicity for all central star-forming galaxies, the parent episodic sample, and the reduced episodic sample used in the stacking analysis. Episodic galaxies account for $\sim 70\%$ of the central SF galaxies in both stellar mass bins. In each mass bin, the stellar mass, SFR, and gas-phase metallicity distributions of the 200 randomly selected galaxies are broadly consistent with those of both the parent episodic sample and the all SF galaxies. The down-selection therefore introduces no evident systematic bias in the galaxy properties most relevant to the subsequent analysis.


\bibliography{EpisodicSFH_FMR}
\bibliographystyle{aasjournal}



\end{document}